# Structural, optical and complex impedance spectroscopy study of multiferroic $Bi_2Fe_4O_9$ ceramic

S. R. Mohapatra[1], B. Sahu[1], T. Badapanda[2], M. S. Pattanaik[1], S. D. Kaushik[3], A. K. Singh[1*]

[1]Department of Physics and Astronomy, National Institute of Technology, Rourkela-769008, Odisha, India.
[2]Department of Physics, C.V. Raman College of Engineering, Bhubaneswar, Odisha, India-752054.
[3] UGC-DAE Consortium for Scientific Research Mumbai Centre, R-5 Shed, BARC, Mumbai-400085, India.

**Abstract:**

Multiferroic bismuth ferrite $Bi_2Fe_4O_9$ (BFO) ceramic was synthesized by conventional solid state reaction route. X-ray diffraction and Rietveld refinement show formation of single phase ceramic with orthorhombic crystal structure (space group *'Pbam')*. The morphological study depicted a well-defined grain of size ~2μm. The optical studies were carried out by using UV-Vis spectrophotometer which shows a band gap of 1.53 eV and a green emission spectrum at 537 is observed in the Photoluminescence study. The frequency dependent dielectric study at various temperature revealed that the dielectric constant decreases with increase in frequency. A noticeable peak shift towards higher frequency with increasing temperature is observed in the frequency dependent dielectric loss plot. The impedance spectroscopy shows a substantial shift in imaginary impedance (Z″) peaks toward the high frequency side described that the conduction in material favoring the long range motion of mobile charge carriers. The presence of non-Debye type multiple relaxations has been confirmed by complex modulus analysis. The frequency dependent ac conductivity at different temperatures indicates that the conduction process is thermally activated. The variation of dc conductivity exhibited a negative temperature coefficient of resistance behavior. The activation energy calculated from impedance, modulus and conductivity data confirmed that the oxygen vacancies play a vital role in the conduction mechanism.

*Keywords: Multiferroic, Dielectric relaxation, Optical properties, Complex Impedance Spectroscopy, Arrhenius behaviour, AC conductivity.*

Email: *singhanil@nitrkl.ac.in*, Phone:+91-661-2462731; Fax:+91-661-2462739

## 1. Introduction:

Multiferroic materials have attracted considerable attention because of their intriguing physical properties, important fundamental issues in spintronics and potential technological applications in multiple-state memories, sensors, filters, actuators and spin devices etc [1]. Ternary bismuth ferrites have received significant interest in the past few years because of their potential applications in actuation, sensing and digital memory [2]. As a typical bismuth ferrite, $Bi_2Fe_4O_9$ [3-6] has been known for several decades, which displays various practical applications such as photocatalyst [7], semiconductor gas sensor [8] and high performance catalyst [9]. We have already reported the bulk $Bi_2Fe_4O_9$ (BFO) ceramics displaying ferroelectric hysteresis loops at T = 250 K and antiferromagnetic ordering ($T_N$ ~ 260 K), indicating that BFO is a promising multiferroic material [10]. Despite the evident importance of $Bi_2Fe_4O_9$ as a functional material, very few reports are associated with its high temperature dielectric measurement and conduction mechanisms. Further, to have a better understanding about its microstructure comprising of grains and grain boundaries, the study of electrical properties are very essential. The complex impedance spectroscopic (CIS) technique is considered to be a promising non-destructive testing method for analyzing the electrical processes occurring in the material on the application of AC signal as input perturbation. Along with, ac conductivity measurement lends precise insight into the contribution of overall electrical conductivity. In this manuscript, we report the detailed structural and electrical study of BFO prepared by solid state reaction route. The structural analysis was done by X-ray diffraction, Rietveld refinement, scanning electron microscopy in parallel with the optical study by UV-Vis spectroscopy, Photoluminiscence study and electrical characterization using complex impedance spectroscopy and conductivity analysis.

## 2. Experimental Technique:

The polycrystalline $Bi_2Fe_4O_9$ was prepared using conventional solid state reaction route. High purity constituent chemicals $Bi_2O_3$ and $Fe_2O_3$ (Aldrich, 99.99%) were taken in stoichiometric ratios. The mixed powders were grounded in an agate motor for 2 h and cylindrical pellets of 1-mm thickness were prepared by using a hydraulic press. The pellets were calcined at 1073 K for 12 h followed by a sintering at 1123 K for 10 h. Room temperature X-ray diffraction (XRD) measurement was performed by the multipurpose X-ray diffraction system, RIGAKU JAPAN/ULTIMA-IV for phase determination. XRD measurement was carried out with Cu-Kα radiation with a step size of $0.002°$ at a slow scan rate of $3°$/min. Surface morphology was studied using Field Emission Scanning Electron Microscope (FESEM), Nova Nano SEM/FEI. In order to study optical properties, UV-Visible spectroscopy was done using Perkin Elmer UV-VIS spectrophotometer (Lambda 35) followed by the photoluminescence (PL) spectra which was measured using a steady state spectrofluorimeter (FLS920-stm), Edinburgh Instruments UK, using 400 nm and 425 nm as an excitation from Xe source. The frequency (0.1 – 1000 kHz) and temperature dependant (323–633 K) dielectric and impedance measurements were carried out using High precession impedance analyser (6500B Wayne Kerr).

## 3. Results and Discussion:
### 3.1. Reitveld Refinement of X-ray diffraction and FESEM analysis:

Fig. 1 shows the Reitveld refined X-ray diffraction patterns of BFO ceramic prepared by solid state reaction route. All diffraction peaks correspond to the orthorhombic mullite-type phase with space group *'Pbam'* which are consistent with the standard data (JCPDS no. 25-0090). The structural refinement was performed using the *FULLPROF* program [11]. The various atomic parameters along with the bond length and bond angles obtained from the reitveld refinement are presented in table 1. Inset of fig. 1(a) depicts the crystal structure of BFO which was drawn by using *DIAMOND 3.2* software. Inset of fig. 1(b) shows the field emission scanning electron micrograph of BFO ceramic which is composed by large grains with an average size of approximately 2μm. It is believed that the microstructural characteristic is related to matter transport mechanism between the grains during the sintering process.

### 3.2. UV-Visible Spectroscopy

Fig. 2 shows the UV–Visible absorbance spectrum of $Bi_2Fe_4O_9$ ceramic. The optical band gap energy ($E_g$) was calculated by an empirical relation proposed by Wood and Tauc [12] given as:

$$\alpha h\nu = A(h\nu - E_g)^n \tag{1}$$

where, $\alpha$ is the optical absorption coefficient, $h\nu$ is the energy (in the units of eV), *A* is the constant (independent of photon energy) and $E_g$ is the optical band gap and n is a constant associated with the different types of electronic transitions (n = 1/2, 2, 3/2 or 3 for direct allowed, indirect allowed, direct forbidden and indirect forbidden transitions respectively). The UV–Vis absorbance spectrum suggests a direct allowed transition and therefore, n = 1/2 was used in the above equation. Fitting in the linear region to $(\alpha h\nu)^2 = 0$ is done and extrapolated to obtain the band gap of BFO which is found to be 1.53 eV. The value of $E_g$ obtained experimentally is very close to the value obtained by theoretical calculations (i.e., $E_g$ = 1.6 eV) [13]. Li et al [14] has also reported that $Bi_2Fe_4O_9$ nanocrystal exhibit band-gap of 1.53 eV. Inset of fig. 2 shows the plot of absorbance spectrum as a function of wavelength for $Bi_2Fe_4O_9$.

### 3.3. Photoluminiscense study:

Fig. 3 shows the room temperature photoluminescence (PL) emission spectrum of BFO at an excitation wavelength of 400 nm. Two distinct emissions were observed in the emission spectrum: one at 484 nm (2.56 eV) and the other at 537 nm (2.31 eV). The first emission corresponds to the blue emission and the lateter corresponds to the green emission. But from fig.3 it is evident that green emission at 537 nm is much stronger than the blue emission at 484 nm. From these observation the smaller emission at 484 nm could be attributed to the presence of defects at grain boundaries. The intense green emission is ascribed to the presence of oxygen vacancies in BFO ceramics. Similar kind of result was also reported by Miriyala et al [15] in $BiFeO_3$ nanotubes. These oxygen vacancies results in strong emission in the PL spectrum and acts as a potential trap for the carriers. It can also be noticed from the inset of fig. 3 that this green emission gets shifted towards red (i.e., from 537 nm to 559 nm) when excitation wavelength is increased from 400 nm to 425 nm. This linear shift can be explained on the basis of transition of photo-induced carriers from oxygen vacancies to the valence band. It results in the green emission (as observed in the present case) thus, changing the intensity and position of the emission spectrum with respect to the excitation energy [16].

### 3.4. Frequency dependence dielectric study:

Fig. 4 illustrates the variation of dielectric permittivity (ε') and dielectric loss (inset) for the BFO ceramic at various temperatures ranging from 323 – 633 K as a function of wide range of frequency. The dielectric constant at low frequency is rather high which is found to decrease with frequency and then becomes more or less stabilized, which corresponds to bulk effect. The frequency dependent tan δ at various temperature shows noticeable peaks at a particular frequency ($f_R$), which increases with increasing temperature as shown in fig. 4 (inset). The shift of the relaxation frequency ($f_R$) with increasing temperature supports both the relaxation scenarios since thermal energy would provide additional driving force to defects dipoles flipping and domain wall mobility [17]. However, the presence of interfacial Maxwell-Wagner (MW) polarization effects could also be the mechanism responsible for the presence of loss peaks. Since, we find relaxation in tan loss is more prominent in the higher temperature range, we have confined our temperature window from 473-633 K for further studies.

### 3.5. Impedance analysis:

Fig. 5(a) shows the frequency dependence of Z′ of BFO for a temperature range of 473 - 633 K. The value of Z′ is higher at lower frequency region (< 20kHz) and as the frequency increases, the value of Z′ decreases monotonically and attains a constant value at the high frequency region for all temperatures, indicates that the conductivity of these compounds increases with the increase in frequency. Decrement of Z′ with the increase of temperature and frequency suggests a possible release of space charge and consequently lowering of the barrier properties of these materials [18]. Frequency dependent imaginary part of impedance (Z″) of BFO ceramic at various temperatures is depicted in fig. 5(b) where the solid lines are fitting to the obtained experimental data. It is done using the equivalent circuit modelled by the Cole-Cole function expressed as [19]:

$$Z^* = \frac{R}{[1+(j\omega\tau)^\alpha]} \qquad (2)$$

where, $\tau = RC$ and $0 \leq \alpha < 1$ implies that the process is governed by a distribution of relaxation times. For an ideal Debye relaxation $\alpha = 1$. The fitted values of α are shown in table 2. The magnitude of Z″ at $f_{max}$ decreases with the temperature indicating the presence of space charge polarization at low frequency which disappears at high frequency [20].

Fig. 6 illustrates the Nyquist plot of BFO at different temperatures between 473 – 633 K. Separation of the grain and grain boundary of the material is obtained by fitting the experimental response to that of an equivalent circuit (shown in inset of fig. 6). The circuit fitting parameter was done by *ZSimpWin* software with the modelled circuits. The experimental value of grain resistance ($R_g$) and grain boundary resistance ($R_{gb}$) at different temperatures have been obtained from the intercept of the semicircular arc on the real axis of impedance (Z′). Both $R_g$ and $R_{gb}$ follow the Arrhenius plot as illustrated in fig. 7. It can be noticed that the activation energy ($E_a$) for ln$R_{gb}$ (0.93 eV) is found to be greater than ln$R_g$ (0.88 eV).

### 3.6. Modulus Analysis:

Fig. 8 (a) and (b) shows the real and imaginary part of electric modulus respectively over a wide range of frequency for a temperature range of 473 – 633 K. The small magnitude of *M′* at low frequency region (<1kHz) confirms negligibly small contribution of electrode effect, but at higher frequency (>1kHz) *M′* value increases gradually. The contribution of short range mobility of charge carriers induces a continuous dispersion at mid frequency region. *M″(ω)* vs *f* plot shows shift in $M″_{max}$ towards higher relaxation frequency with increase in temperature. It confirms that relaxation process is thermally activated process where a hopping mechanism of charge carriers dominates mostly at higher temperature. The solid lines in fig. 8(b) are fitted

curves according to Bergman modified KWW (Kohlraush-Williams-Watts) function [21] which is given as:

$$M''(\omega) = \frac{M''_{max}}{\{1-\beta+(\frac{\beta}{1+\beta})[\beta(\omega_{max}/\omega)+(\omega/\omega_{max})]^{\beta}\}} \quad (3)$$

where, $M''_{max}$ is the peak maximum of $M''(\omega)$, $\omega_{max}= 1/\tau$ ($\tau$ is conductivity relaxation time) is the peak frequency of $M''(\omega)$, the exponent $\beta$ symbolises the degree of non-Debye behaviour and is related to full width half maxima of $M''(\omega)$ vs $f$ curve. Inset of fig.8(b) shows the variation of $\beta$ with temperature. For $\beta = 1$ relates to ideal Debye nature with unique relaxation time and $\beta = 0$ denotes maximum interaction of dipoles with other dipoles. Fig.9 depicts the scaling behaviour of BFO where we have plotted the normalised modulus parameters, i.e., $M''/M''_{max}$ and $Z''/Z''_{max}$ versus $f/f''_{max}$. Here, $f''_{max}$ is the value corresponding to the maximum value of $M''_{max}$ and $Z''_{max}$ at different selected temperatures (473–593 K). We observe that all the peaks for different temperature coincides which reveals temperature-independent behaviour of the dynamic process taking place in the material. The overlapping peaks reveal the evidence of transition from long-range to short-range mobility with an increase in frequencies. The imaginary component of impedance (Z'') and modulus (M'') as a function of frequencies at selected temperatures (i.e. 493 K, 533 K and 573 K) are plotted in fig. 10(a-c). Sinclair et al [22] suggested that the presence of smallest capacitance and largest resistance can be well understood using the frequency dependent combined plots of Z'' and M''. It helps to identify whether there is long-range or short-range motion of charge carriers. Fig. 10 (a-c) clearly depicts the mismatch of peaks corresponding to Z'' and M'' at different frequencies for all temperatures. The mismatch of the peaks of Z'' and M'' confirms the short-range motion of charge carriers. Fig. 11 shows the $ln\,\tau_m$ versus $1000/T$ plot where frequency ($\omega_m = 1/\tau_m$) corresponds to the peak position of $Z''_{max}$ and $M''_{max}$. The most probable relaxation frequency is determined from the relation $\omega_m \tau_m = 1$ where $\tau_m$ is the most probable relaxation time [23].The linear fitting of the experimental data points are done and it is found that $\tau_m$ follows the Arrhenius law written as:

$$\tau_m = \tau_0 \exp(-E_a/k_B T) \quad (4)$$

where, $\tau_0$ is pre-exponential factor, $E_a$ is activation energy and $k_B$ is the Boltzmann constant. The $E_a$ was calculated to be 0.91 eV and 0.87 eV for the relaxation of $Z''_{max}$ and $M''_{max}$ respectively.

### 3.7. AC conductivity study:

The ac electrical conductivity is derived from the impedance data using the relation [24]:

$$\sigma_{ac} = \frac{Z'}{[(Z')^2 + (Z'')^2]} \times \frac{t}{A} \tag{5}$$

where, t and A are the thickness and surface area of the material respectively. Fig. 12 shows the variation of electrical conductivity ($\sigma_{ac}$) with frequency at different temperatures ranging from 473 – 533 K. The frequency independent plateau at a low frequency is attributed to the long-range translational motion of ions contributing to dc conductivity ($\sigma_{dc}$) [25]. A convenient formalism to investigate the frequency behavior of conductivity at constant temperature in a variety of materials is based on the power law proposed by Jonscher [26], $\sigma_{ac} = \sigma_{dc} + A\omega^n$, where, A is pre-exponential constant, $\omega$ (= $2\pi f$) is the angular frequency and n is the frequency exponent (0<n<1). The values of A and n obtained from single power law fitting is listed in table 2. The variation of $\sigma_{dc}$ against $10^3/T$ is plotted in the inset of fig. 12. The value of bulk conductivity of the material at different temperatures is evaluated from the ac conductivity plot of the sample by theoretical fitting using Joncher's power law [27]. At higher temperature, the conductivity versus temperature response is more or less a linear response and can be explained by a thermally activated transport of Arrhenius type governed by the relation:

$$\sigma_{dc} = \sigma_0 \exp(-E_a/k_B T) \tag{6}$$

where $\sigma_o$, $E_a$ and $k_B$ represent the pre-exponential term, the activation energy of the mobile charge carriers and Boltzmann's constant respectively. The $\sigma_{dc}$ activation energy ($E_a$) of the material is calculated to be 0.86 eV. The calculated $E_a$ values for BFO ceramic using different parameters are found to lie in the range of 0.86–0.93 eV, indicating that its electrical conduction is largely dominated by the same thermal excitation of carriers from the short-range motion of oxygen vacancies. Moreover, the activation energy values calculated from the relaxation are very close to the values calculated from the conduction process, confirming that the same type of charge carriers (i.e., oxygen vacancies) should be responsible for both dielectric relaxation and electrical conduction.

## 4. Conclusion:

Single phase $Bi_2Fe_4O_9$ ceramic was synthesised by solid state route. Reitevld refinement of XRD shows that the phase pure material has an orthorhombic crystal structure with space group *'Pbam'*. The band gap of the ceramic was found to be 1.54 eV and a green emission spectra at 537 nm from PL study attributed to oxygen vacancies in BFO. The CIS study revealed that the electrical relaxation process was temperature dependent and non-Debye type. The compound exhibited NTCR behavior and temperature dependent relaxation phenomena. The ac conductivity spectrum was found to obey Jonscher's universal power law and increased with temperature. Finally, the activation energy ($E_a$) obtained from impedance, modulus spectroscopic curve and dc conductivity indicated that oxygen ion vacancies led to the conduction mechanism at higher temperature.

**Acknowledgements**

We acknowledge Board of Research in Nuclear Science (BRNS), Mumbai (Sanction No: 2012/37P/40/BRNS/2145) and Department of Science and Technology (DST), New Delhi (Sanction No: SR/FTP/PS-187/2011) for funding.SRM and BS acknowledge BRNS and DST, India respectively for the financial support. Lastly, SRM is thankful to Dr. P. K. Sahoo, NISER (BBSR) for PL characterization and Tapabrata Dam for his useful suggestions.

**References**

1. N.A. Spaldin, S. Cheong, R. Ramesh, Multiferroics: Past,present, and future. Physics Today **63**, 10, 38-43 (2010)
2. Y. Liu, R. Zuo, Morphology and optical absorption of $Bi_2Fe_4O_9$ crystals via mineralizer-assisted hydrothermal synthesis. Particuology **11**, 581–587 (2013)
3. M. N. Iliev, A. P. Litvinchuk, V. G. Hadjiev, M. M. Gospodinov, V. Skumryev, and E. Ressouche, Phonon and magnon scattering of antiferromagnetic Bi2Fe4O9. Phys. Rev. B **81**, 024302 (2010)
4. D.H. Wang, C.K. Ong,The phase formation and magnetodielectric property in (1−x)$Bi_2Fe_4O_9$–xBaO composites. J. Appl. Phys. **100**, 044111 (2006)
5. Tae-Jin Park, G.C. Papaefthymiou, A. R. Moodenbaugh, Y. Mao, S.S. Wong, Synthesis and characterization of submicron single-crystalline $Bi_2Fe_4O_9$cubes. J. Mater. Chem.,**15**, 2099–2105 (2005)

6. Jian-Tao Han, Yun-Hui Huang, Rui-JieJia, Guang-Cun Shan, Rui-Qian Guo, W. Huang, Synthesis and magnetic property of submicron $Bi_2Fe_4O_9$. J. Cryst. Growth **294**, 469–473 (2006)

7. Q. Zhang, W.J. Gong, J.H. Gong, X.K. Ning, Z.H. Wang, X.G.Zhao et al., Size-dependant magnetic, photoabsorbing and photocatalytic properties of single-crystalline $Bi_2Fe_4O_9$ semiconductor nanocrystals. J. Phys. Chem. C **115**, 25241-25246 (2011)

8. A.S. Poghossian, H.V. Abobian, P.B. Avakian et al., Bismuth ferrites: New materials for semiconductor gas sensors. Sens. Actuators B Chemical **4**, 545-549 (1991)

9. S.M. Sun, W.Z. Wang, L. Zhang, M. Shang, Visible Light-Induced Photocatalytic Oxidation of Phenol and Aqueous Ammonia in Flowerlike $Bi_2Fe_4O_9$ Suspensions. J. Phys. Chem. C **113**, 12826-12831 (2009)

10. A.K. Singh, S.D. Kaushik, B. Kumar, P.K. Mishra, A. Venimadhav, V. Siruguri, S. Patnaik, Substantial magnetoelectric coupling near room temperature in $Bi_2Fe_4O_9$. Appl. Phys. Lett. **92**, 132910 (2008)

11. H.M. Rietveld, A profile refinement method for nuclear and magnetic structures. J. Appl. Cryst. **22**, 65-71 (1969)

12. D.L. Wood, J. Tauc, Weak Absorption Tails in Amorphous Semiconductors. Phys. Rev. B. **5**, 3144 -3151 (1972)

13. Z. Irshad, S.H. Shah, M.A. Rafiq, M.M. Hasan, First principles study of structural, electronic and magnetic properties of ferromagnetic $Bi_2Fe_4O_9$. J. Alloys Compd. **624,** 131–136 (2015)

14. Y. Li, Y. Zhang, W. Le, J. Yu, C. Lu, L.Xia, Photo-to-current response of $Bi_2Fe_4O_9$ nanocrystals synthesized through a chemical co-precipitation process. New J. Chem., **36**, 1297–1300 (2012)

15. N. Miriyala, K. Prashanthi, T. Thundat, Oxygen vacancy dominant strong visible photoluminescence from $BiFeO_3$ nanotubes. Phys. Status Solidi RRL **7**, 668–671 (2013)

16. I. Mora-Sero, J. Bisquert, Fermi Level of Surface States in $TiO_2$ Nanoparticles. Nano Lett. **3**, 945 (2003)

17. A. Dutta, T.P. Sinha, Dielectric relaxation and electronic structure of $Ca(Fe_{1/2}Sb_{1/2})O_3$ Phys. Rev. B**46,** 155113 (2007)

18. V.S. Postnikov, V.S. Pavlov, S.K. Turkov, Internal friction in ferroelectrics due to interaction of domain boundaries and point defects. J. Phys. Chem. Solids **31**, 1785-1791 (1970)
19. K.S. Cole, R.H. Cole, Dispersion and Absorption in Dielectrics I. Alternating Current Characteristics. J. Chem. Phys. **9**, 341 (1941)
20. S. Sen, S.K. Mishra, S.K. Das, A. Tarafdar, Impedance analysis of $0.65Pb(Mg_{1/3}Nb_{2/3})O_3$–$0.35PbTiO_3$ ceramic. J. Alloys Compd. **453**, 395-400 (2008)
21. T. Badapanda, S. Sarangi, S. Parida, B. Behera, B. Ojha, S. Anwar, Frequency and temperature dependence dielectric study of strontium modified Barium Zirconium Titanate ceramics obtained by mechanochemical synthesis. J. Mater. Sci: Mater Electron. **26**, 3069-3082 (2015)
22. R. Kohlrausch, Theorie des elektrischen Rückstandes in der Leidener Flasche. Pogg. Ann. Phys. Chem. **91**, 179 (1854)
23. D.C. Sinclair, A.R. West, Impedance and modulus spectroscopy of semiconducting $BaTiO_3$ showing positive temperature coefficient of resistance. J. Appl. Phys. **66**, 3850 (1989)
24. M.M. Hoque, A. Dutta, S. Kumar, T.P. Sinha, Dielectric relaxation and conductivity of $Ba(Mg_{1/3}Ta_{2/3})O_3$ and $Ba(Zn_{1/3}Ta_{2/3})O_3$. J. Mater. Sci. Technol. **30**, 311-320 (2014)
25. W. Li, R.W. Schwartz, ac conductivity relaxation processes in $CaCu_3Ti_4O_{12}$ ceramics: Grain boundary and domain boundary effects. Appl. Phys. Lett. **89**, 242906 (2006)
26. K. Funke, Jump relaxation in solid electrolytes. Prog. Solid State Chem. **22**, 111–195 (1993)
27. A.K. Jonscher, The `universal' dielectric response. Nature **267**, 673-679 (1977)

**Figure Captions:**

Figure 1: Rietveld refinement of XRD data of $Bi_2Fe_4O_9$ ceramic. Solid hollow circle, red line, blue line and green bar mark represent observed, calculated, difference between observed and calculated and Bragg positions respectively. Inset shows crystal structure of BFO representing octahedral (O) and tetrahedral (T) coordinated Fe atoms with Oxygen atoms and FESEM micrographs of BFO.

Figure 2: $(\alpha h\upsilon)^2$ versus energy (eV) plot displaying the energy band gap of $Bi_2Fe_4O_9$. Inset shows the absorption spectra plot as a function of wavelength.

Figure 3: Room temperature PL spectrum of $Bi_2Fe_4O_9$ at an excitation wavelength of 400 nm. Inset shows shift in emission spectrum obtained for an emission wavelength of 400 nm and 425 nm.

Figure 4: Frequency dependant dielectric permittivity ($\varepsilon'$) and tan $\delta$ (inset) at various temperatures ranging from 323-633 K

Figure 5: (a) Real part and (b) imaginary part of impedance spectrum as a function of frequency at different temperatures for BFO ceramic.

Figure 6: Nyquist plot for BFO ceramic at different temperatures (solid dots represents experimental data points and solid lines shows theoretical fit). Inset shows Nyquist plot for temperature range 473-553 K and the equivalent circuit.

Figure 7: Arrhenius plot of resistance for grains ($R_g$) and grain boundaries ($R_{gb}$) of BFO ceramic.

Figure 8: (a) Real and (b) imaginary part of electric modulus as a function of frequency at different temperatures for BFO ceramic.

Figure 9: Scaling behaviour of imaginary part of electric modulus and impedance as a function of frequency at selected temperatures (473, 513, 553and 593 K).

Figure 10: The imaginary component of impedance (Z") and modulus (M") as a function of various frequencies at selected temperatures (493, 533 and 573 K).

Figure 11: Relaxation behaviour (Arrhenius plot) of maximum frequency from imaginary part of impedance and electric modulus spectrum (relaxation time $\tau = 1/\omega$) for BFO ceramic.

Figure 12: Frequency dependence of ac conductivity ($\sigma_{ac}$) at selected temperatures (473 K-533 K) for BFO ceramic. Inset shows temperature dependence of dc conductivity ($\sigma_{dc}$) obeying Arrhenius behaviour for BFO ceramic.

**Table 1**         **Crystal structure parameters of Bi$_2$Fe$_4$O$_9$**

| Parameters | | | |
|---|---|---|---|
| Crystal system | Orthorhombic | | |
| Space group | Pbam (No. 55) | | |
| Cell volume (Å$^3$) | 403.94 | | |
| Lattice Parameter | | | |
| a (Å) | 7.9730 | | |
| b (Å) | 8.4410 | | |
| c (Å) | 6.0020 | | |
| Atomic Positions | x | y | z |
| Bi-4g | 0.1761 | 0.1715 | 0.0000 |
| Fe(1)-4f | 0.5000 | 0.0000 | 0.2570 |
| Fe(2)-4h | 0.3510 | 0.3340 | 0.5000 |
| O(1)-2b | 0.0000 | 0.0000 | 0.5000 |
| O(2)-8i | 0.3730 | 0.1980 | 0.2430 |
| O(3)-4h | 0.1370 | 0.4180 | 0.5000 |
| O(4)-4g | 0.1430 | 0.4150 | 0.0000 |
| Discrepancy factor | | | |
| R$_p$ (%) | 13.2 | | |
| R$_{wp}$ (%) | 15.1 | | |
| R$_{exp}$(%) | 7.89 | | |
| $\chi^2$ | 3.66 | | |
| Bond Length | | | |
| Bi-O(2) | 2.154 Å | | |
| Bi-O(4) | 2.072 Å | | |
| Fe(1)-O(2) | 1.955 Å | | |
| Fe(1)-O(3) | 1.949 Å | | |
| Fe(1)-O(4) | 2.047 Å | | |
| Fe(2)-O(1) | 1.837 Å | | |
| Fe(2)-O(2) | 1.930 Å | | |
| Fe(2)-O(3) | 1.847 Å | | |
| Bond Angle | | | |
| O(2)-Bi-O(2) | 85.21$^0$ | | |
| O(4)-Bi-O(2) | 89.41$^0$ | | |
| O(2)-Fe(1)-O(3) | 92.6$^0$ | | |
| O(2)-Fe(1)-O(4) | 87.5$^0$ | | |
| O(3)-Fe(1)-O(4) | 97.3$^0$ | | |
| O(4)-Fe(1)-O(4) | 82.2$^0$ | | |
| O(2)-Fe(2)-O(1) | 113.2$^0$ | | |
| O(2)-Fe(2)-O(2) | 106.0$^0$ | | |
| O(2)-Fe(2)-O(3) | 108.1$^0$ | | |
| O(3)-Fe(2)-O(1) | 107.7$^0$ | | |

**Table 2**

**Values of different parameters obtained from Z" and ac conductivity:**

| Temp (K) | Z" | ac conductivity | |
|---|---|---|---|
| | $\alpha$ | A | n |
| 473 | 0.89 | $8.29\times10^{-12}$ | 0.74 |
| 493 | 0.89 | $1.06\times10^{-11}$ | 0.72 |
| 513 | 0.88 | $1.14\times10^{-11}$ | 0.71 |
| 533 | 0.87 | $2.24\times10^{-11}$ | 0.69 |
| 553 | 0.86 | $4.48\times10^{-11}$ | 0.68 |
| 573 | 0.85 | $1.03\times10^{-10}$ | 0.66 |
| 593 | 0.84 | $2.77\times10^{-10}$ | 0.63 |
| 613 | 0.83 | $4.86\times10^{-10}$ | 0.62 |
| 633 | 0.82 | $5.84\times10^{-10}$ | 0.60 |

**Figure 1**

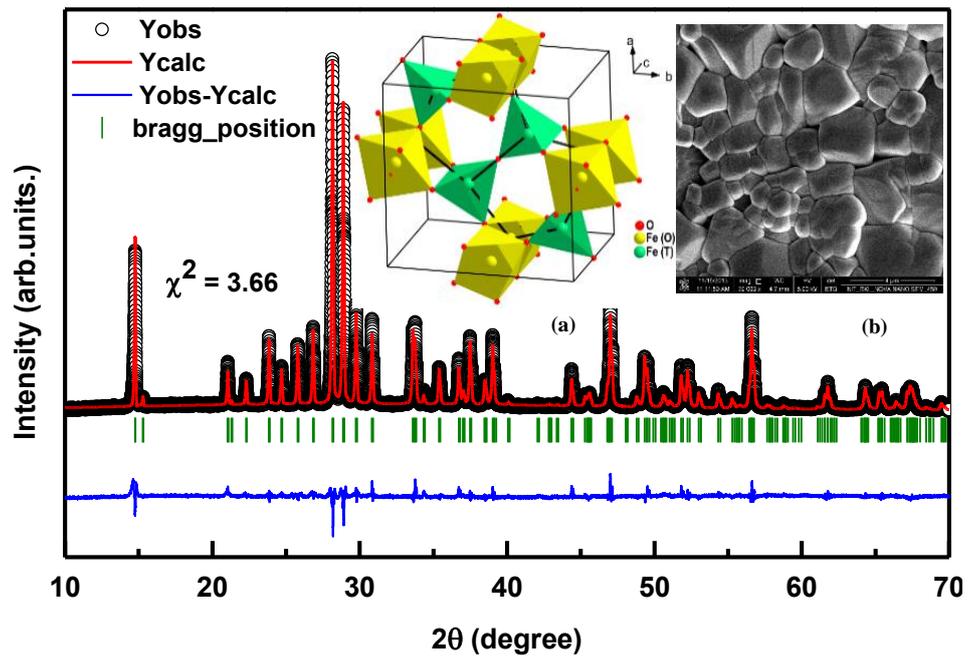

**Figure 2**

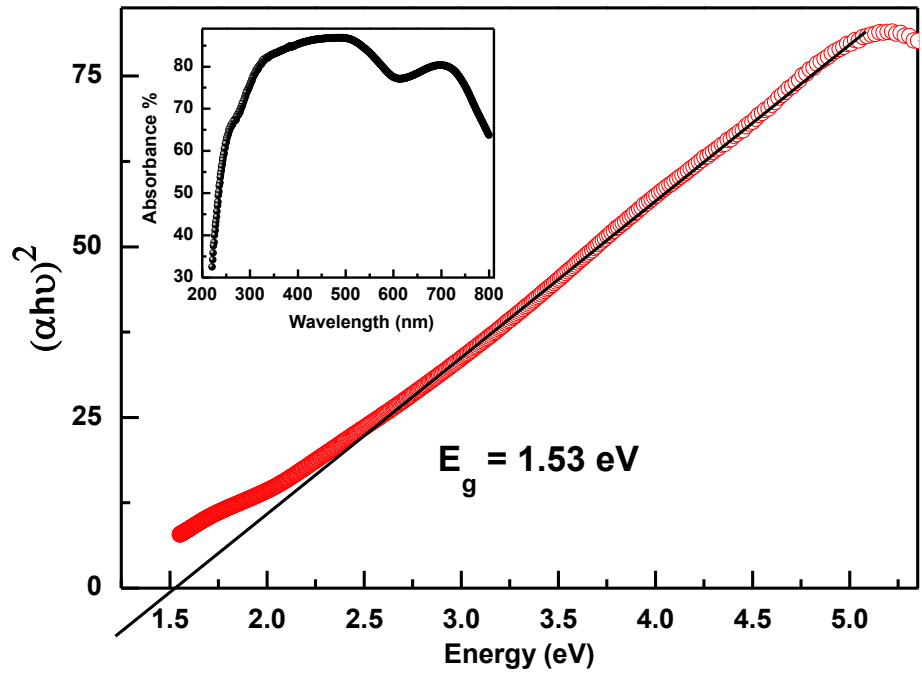

**Figure 3**

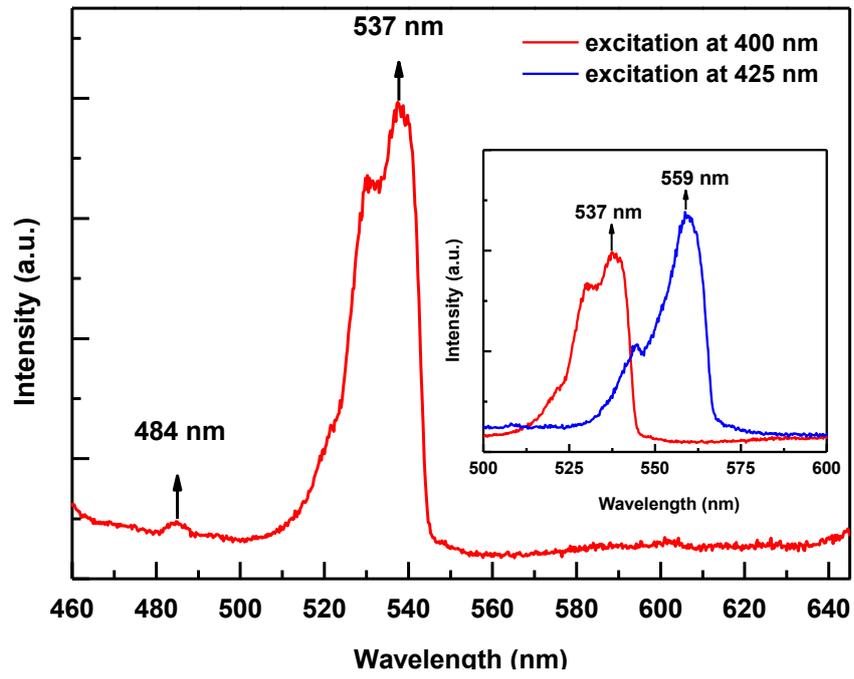

**Figure 4**

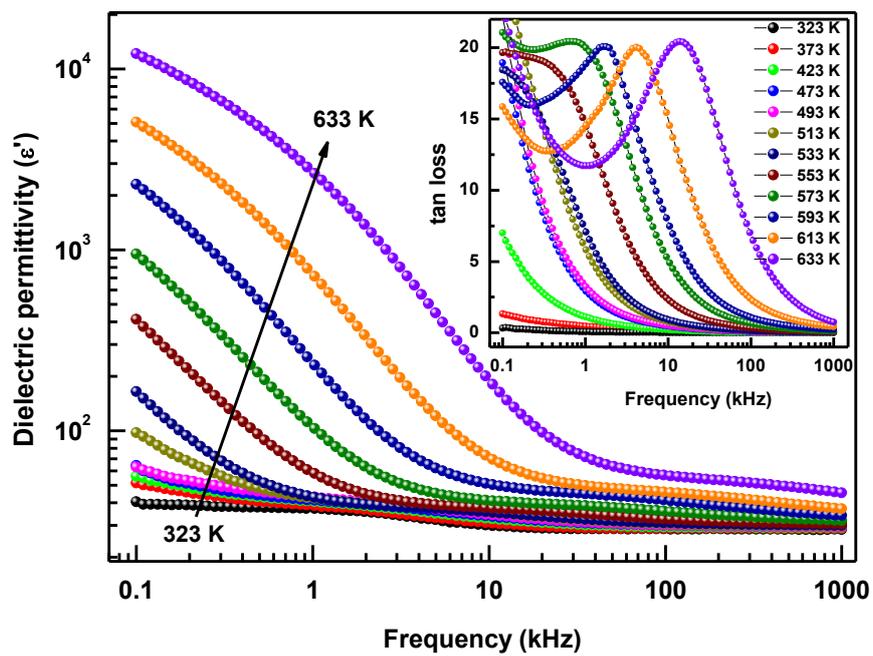

**Figure 5**

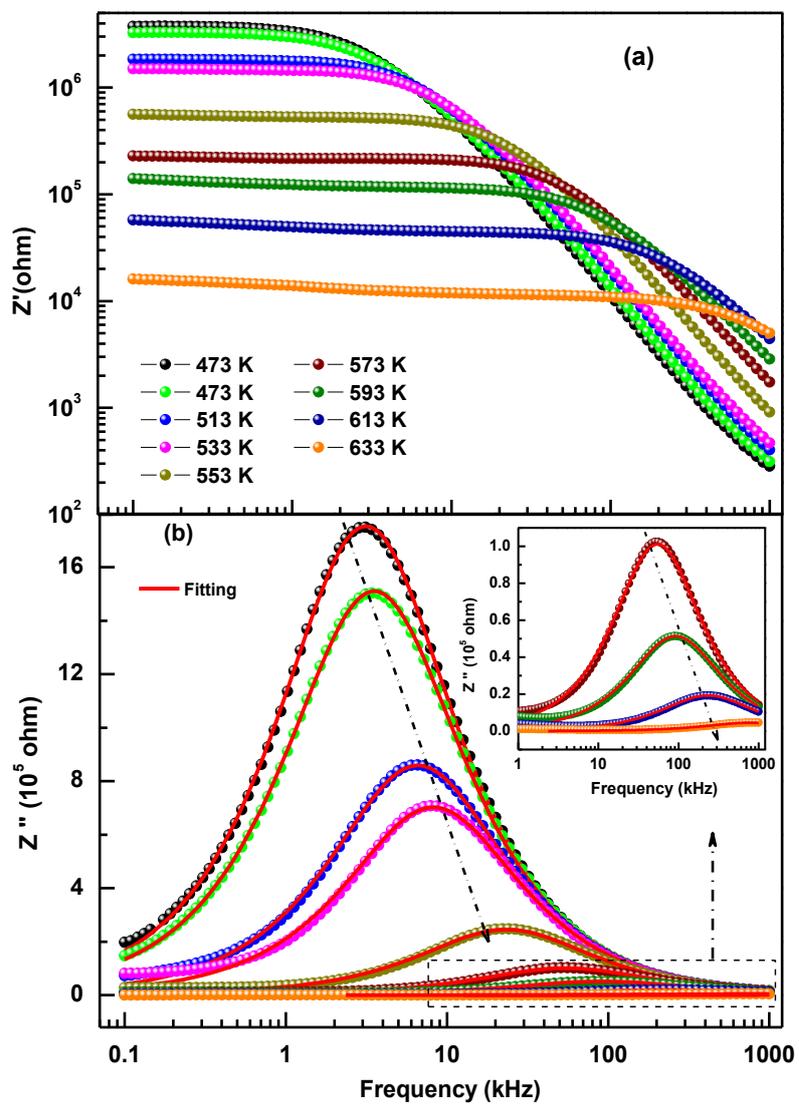

**Figure 6**

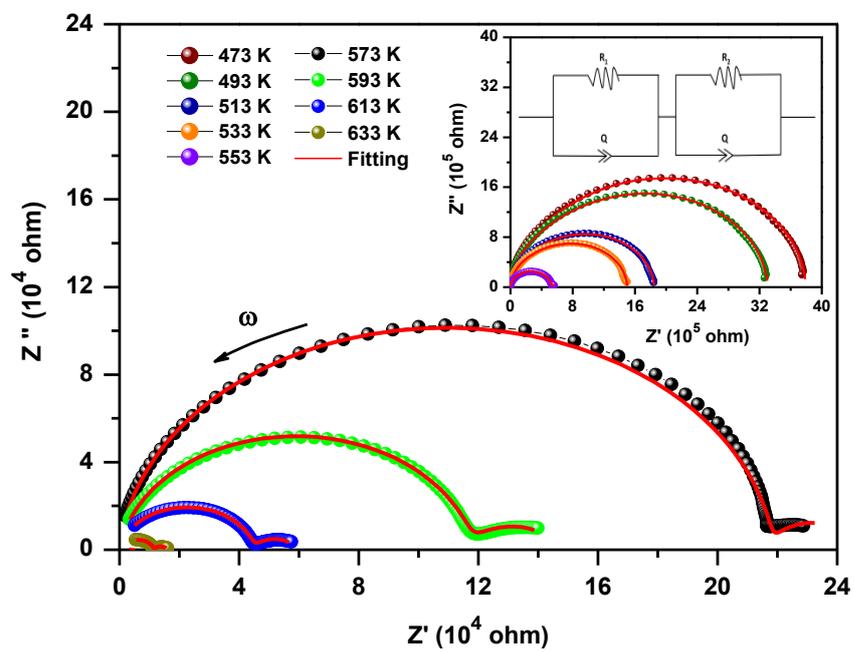

**Figure 7**

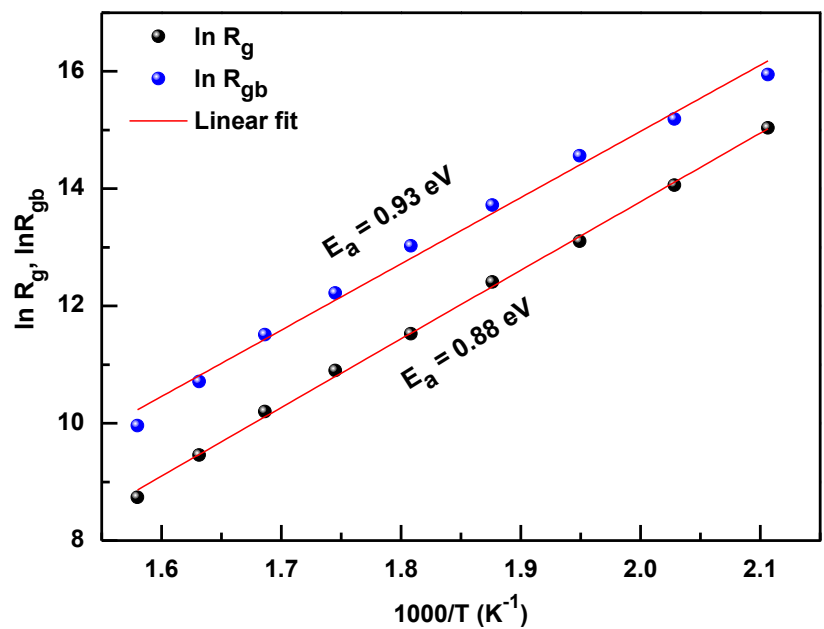

**Figure 8**

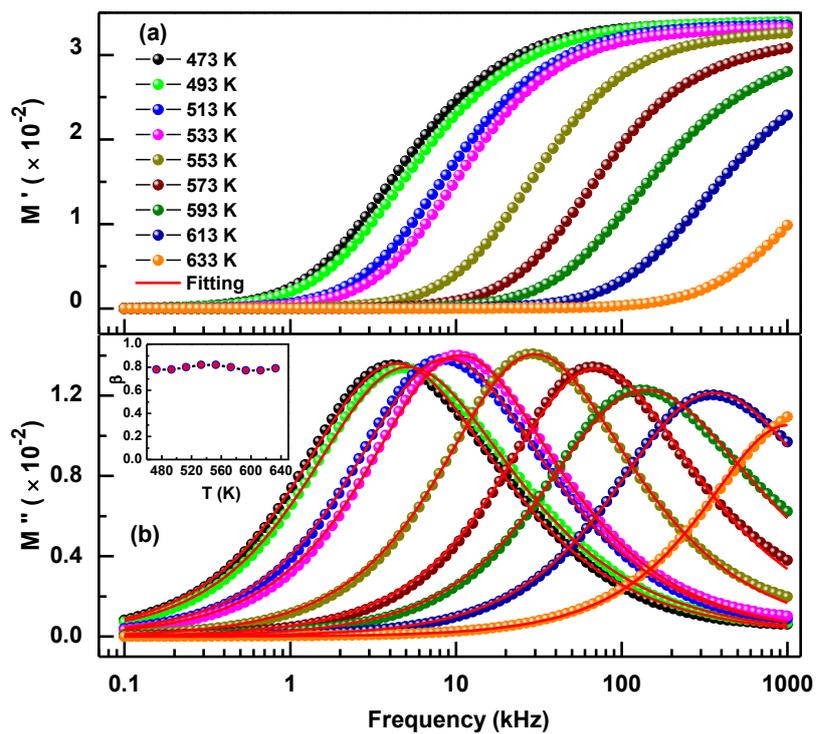

**Figure 9**

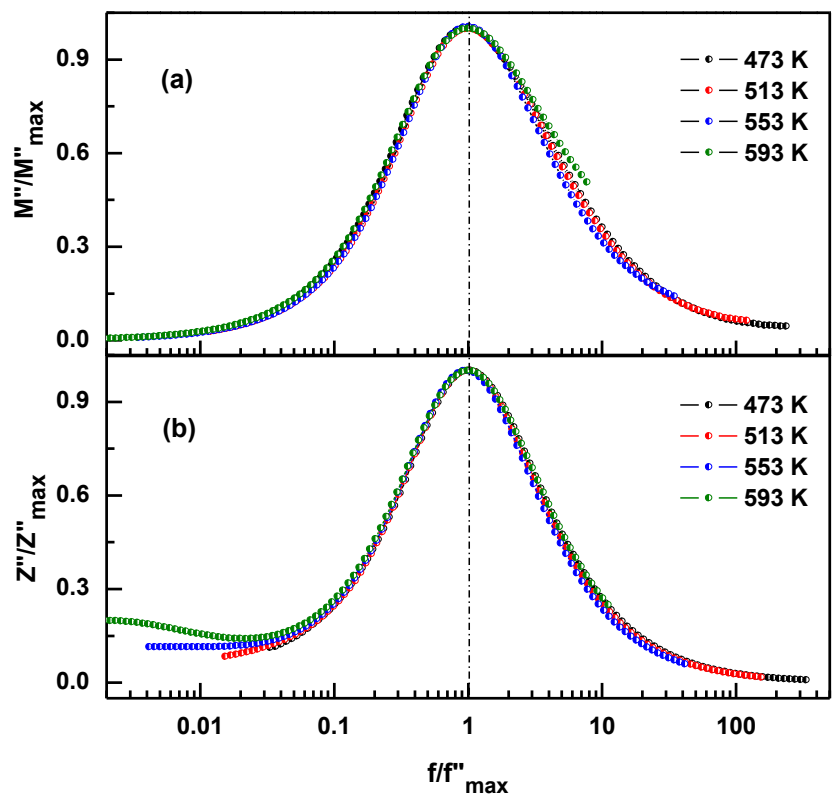

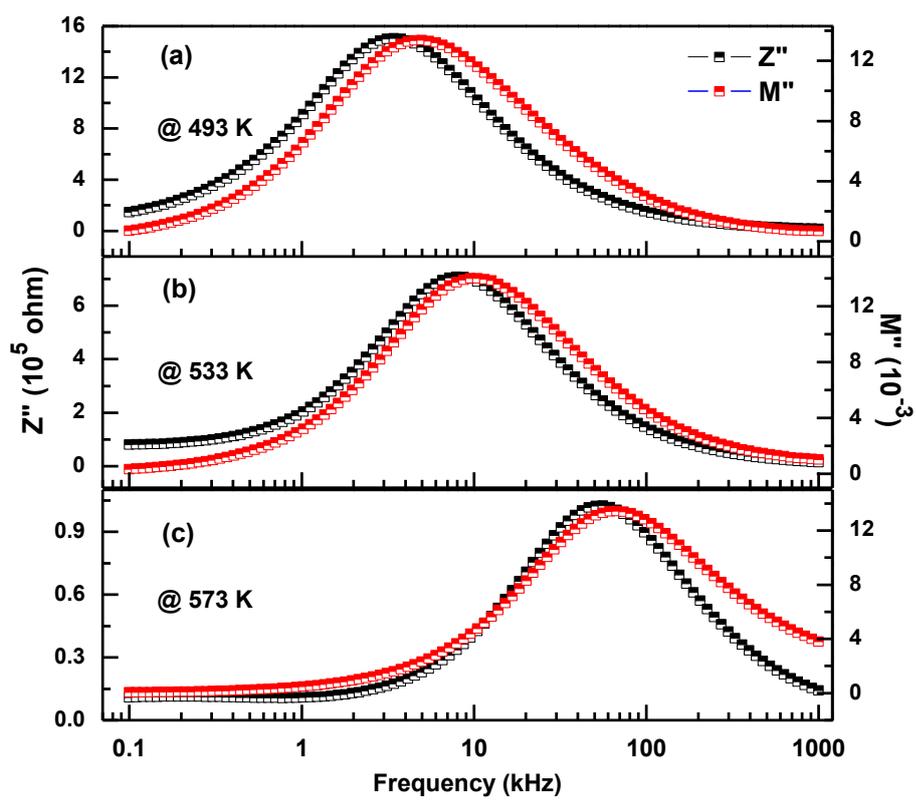

**Figure 10**

**Figure 11**

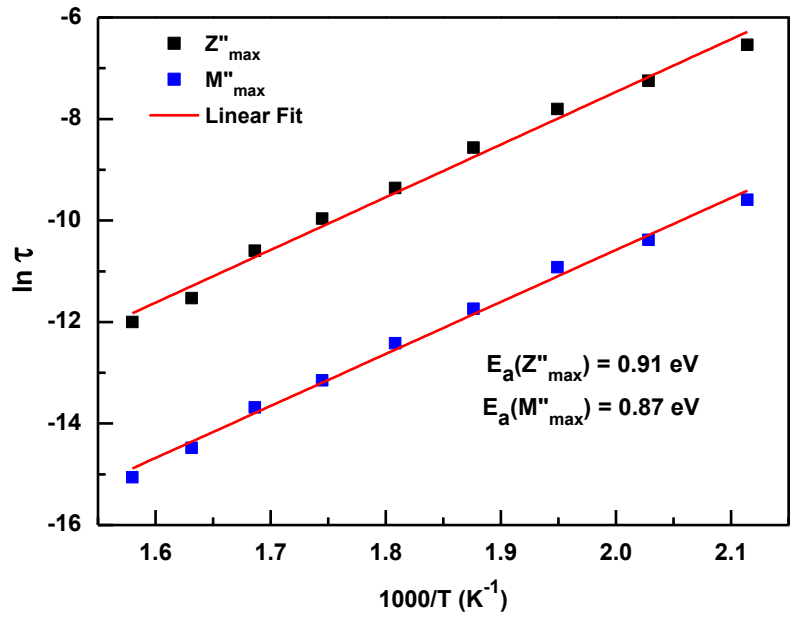

**Figure 12**

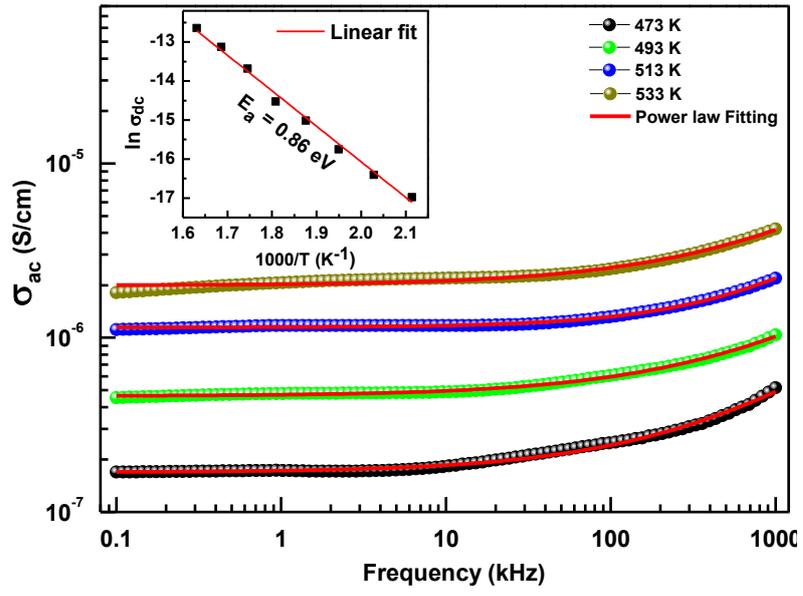